\documentclass{article}
\usepackage{epsfig}
\begin{document}
\begin{flushright}
{\bf {DFUB 15/2002 \\ Bologna 12/11/2002 }}
\end{flushright}
\vspace{0.5cm}
\begin{center}
{\Large{\bf {MACRO results on atmospheric neutrino oscillations}}} \par
\vspace{1cm}
Y.~Becherini for the MACRO collaboration \\ 
Physics Department, University of Bologna and INFN, \\
viale C. Berti Pichat 6/2, \\ 
I-40127 Bologna, Italy \\ 
Yvonne.Becherini@bo.infn.it \par
\vspace{0.5cm}
Proceedings for the International School of Physics ``Enrico Fermi'', \\
CLII course ``Neutrino Physics'', \\
Varenna, Italy, 23 July - 2 August  2002.

\end{center}

\begin{abstract}
In this paper I shall resume the results of the MACRO experiment on 
atmospheric muon neutrino oscillations.  
\end{abstract}

\section{Atmospheric neutrinos and MACRO}
A high energy primary cosmic ray interacts
in the upper atmosphere producing a large number of pions
and kaons, which decay yielding muons and muon
neutrinos; the muons decay generating muon and electron neutrinos.
These neutrinos are produced in a spherical shell at $10-20$ km 
above ground and travel towards the Earth. 
The $\nu_{\mu}$ can be detected by an underground experiment as muons, 
after their charged current interactions with the material inside or outside the detector 
($\nu_{\mu} N \rightarrow \mu^{+-} +...$).

MACRO was a large area multipurpose underground experiment optimized to search 
for rare events in the cosmic radiation.
The detector was located in the Gran Sasso laboratory, 
which is well shielded from downgoing cosmic ray muons by a mean 
rock thickness of 3700 m.w.e.. 
MACRO was composed of three sub-detectors: liquid scintillation counters for energy and T.O.F. 
measurements, limited streamer tubes for particle tracking and nuclear track detectors 
for magnetic monopoles and nuclearites searches. 
It had a modular structure: it was divided into six sections referred to as supermodules, 
each of which had a size of $12.6 \times 12 \times 9.3$ m$^3$. 
A cross section of the detector is shown in Fig. 1 on the left, with the different event
topologies of $\nu_\mu$ interactions in or around the detector: {\it Internal Upgoing} 
(IU), {\it Upgoing Stopping} (UGS) and {\it Internal Downgoing} (ID), 
and {\it Up throughgoing} muons.   
Data were taken from March 1989 till April 1994 with the detector under 
construction and from the middle of 1994 till the end of 2000 with the full detector. \par

\section{Neutrino oscillations analysis}
{\bf{Neutrino oscillations.}}
In the simple case of only two neutrinos with flavour eigenstates
$(\nu_\mu,~\nu_\tau)$ the survival probability of a pure $\nu_\mu$ beam is:
\begin{equation}
P(\nu_\mu \rightarrow \nu_\mu) = 1 - P(\nu_\mu \rightarrow \nu_\tau) =
1 - \sin^2 2\theta_{23}~\sin^2 \left( {{1.27 \Delta m^2 \cdot L}\over {E_\nu}} \right)
\end{equation}

\noindent where $P(\nu_\mu \rightarrow \nu_\tau)$ is the oscillation probability, 
$\theta_{23}$ is the mixing angle, ($\nu_2,\nu_3$) are the mass eigenstates, 
$\Delta m^2=m^2_3-m^2_2$ (eV$^2$), $L$ (km) is the distance travelled by the neutrino 
from the production point to the detection point, and $E_\nu$ (GeV) the neutrino energy. \par

\begin{figure}
\begin{center}
\mbox{
\includegraphics[width=6.3cm]{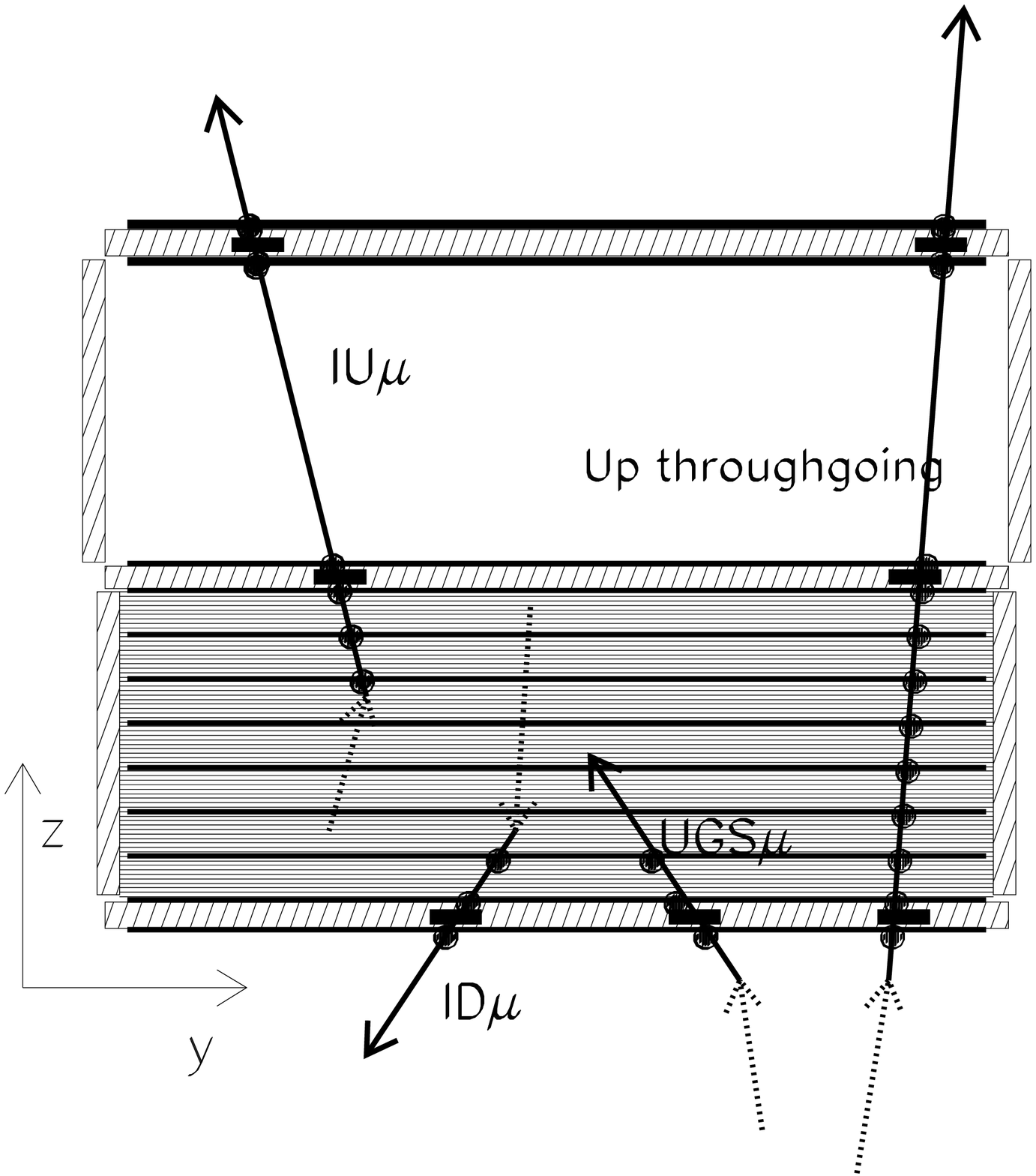}
\hspace{0.3cm} \vspace{-1cm}
\includegraphics[width=6.3cm]{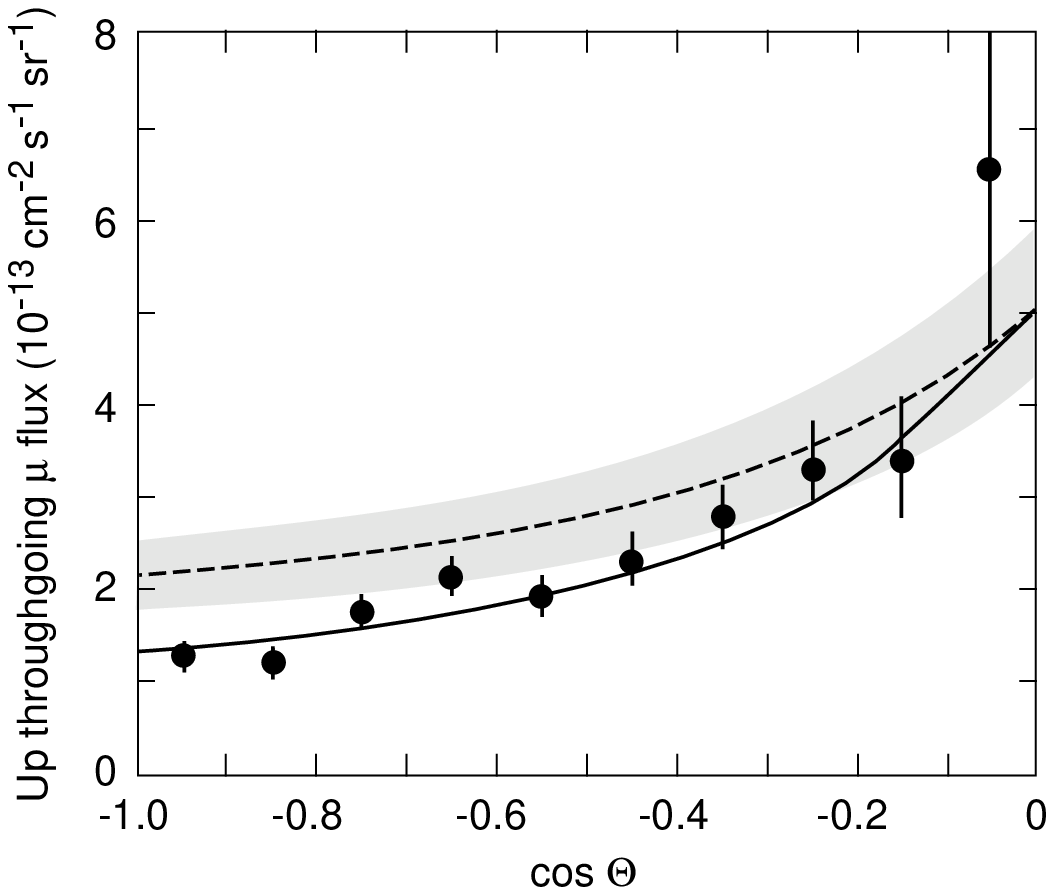}}
\end{center}
\caption{\label{fig1} \small 
{\bf{Left:}} Cross section of the MACRO detector and sketch of different event
topologies induced by $\nu_\mu \rightarrow \mu $ charged current interactions.
The black points and the black rectangles represent streamer tubes and
scintillator hits, respectively. 
{\bf{Right:}} Zenith angle distribution of up throughgoing muons (black points).
The dashed line is the expectation for no oscillations 
(with a 17 \% scale uncertainty band). The solid line is the fit
for an oscillated muon flux which yielded maximum mixing and 
$\Delta m^{2} =2.5\cdot 10^{-3}$ eV$^{2}$ [2].}
\end{figure}

{\bf{Up throughgoing muons.}} The up throughgoing muons come from 
$\nu_{\mu}$ interactions below the detector. 
These neutrinos have an average energy of $\sim 50$ GeV \cite{macro1},\cite{macro2}. 
The data (6.16 years of livetime) deviate in absolute value and in zenith angle distribution 
from Monte Carlo (MC) predictions without oscillations, see Fig. 1 on the right. 
Fig. 2 upper left shows the allowed region for \( \nu _{\mu }\rightarrow \nu _{\tau } \) 
oscillations in the \( sin^{2}2\theta - \Delta m^{2}\) plane, computed according to \cite{macro3} 
for the up throughgoing muon events;
our allowed region is compared with those obtained by the SuperKamiokande (SK) \cite{macro3} 
and Soudan 2 \cite{macro4} experiments. 
\par
Matter effects due to the difference between the weak interaction
effective potential for muon neutrinos with respect to sterile neutrinos would produce 
a different total number and a different zenith distribution of up throughgoing muons \cite{macro6}.
In Fig. 2 on the upper right the measured ratio between the nearly vertical events with
$ -1 < cos \Theta < -0.7$ and the horizontal events with $-0.4 < cos \Theta < 0$ is shown as a black point. 
In this measurement most of the theoretical uncertainties on neutrino flux and cross section cancel. 
The remaining theoretical error combined with the experimental error is estimated to be $7\%$. 
We measured 305 events with $ -1 < cos \Theta < -0.7$ and 206 with $-0.4 < cos \Theta < 0 $;
the ratio is R = $1.48 \pm 0.13_{stat} \pm 0.10_{sys}$.
For $ \Delta m^{2}=2.5\cdot 10^{-3} eV ^{2} $ and maximal
mixing, the expected value of the ratio for $ \nu _{\mu }\rightarrow \nu _{\tau } $
is $R_{\tau} = 1.72$ while for $ \nu _{\mu }\rightarrow \nu _{s} >$ is $R_{sterile} = 2.16$.
$ \nu _{\mu }\rightarrow \nu _{s} $ oscillations are disfavoured at 99\% c.l. compared 
to the $ \nu _{\mu }\rightarrow \nu _{\tau } $ channel \cite{macro2}, \cite{macro5}. 

The oscillation probability is a function of the ratio $L/E_\nu$, see Eq. 1. 
In MACRO the neutrino energy was estimated for each event by measuring the muon energy 
$E_\mu$, via Multiple Coulomb Scattering (MCS) of the induced muon in the horizontal absorbers 
in the lower MACRO (Fig. 1, left).
The r.m.s. of the lateral displacement for a muon crossing the 
apparatus on the vertical is $\sigma_{MCS}\simeq 10$ cm / $E_{\mu}$ (GeV). 
A first analysis was made studying the deflection of up throughgoing
muons with the streamer tubes in \lq\lq digital mode''. 
This method could reach a spatial resolution of $\sim 1~cm$ which implies a maximum 
measurable energy of 10 GeV \cite{macro6}.
To improve the spatial resolution of the detector, 
a second analysis was performed with the streamer tubes used in {}``drift mode{}''
and the resolution achieved was $ \simeq 3~mm $. 
The method allowed the separation of the up throughgoing
muons in 4 subsamples corresponding to average neutrino energies  of 12, 20, 
50 and 102 GeV, respectively. 
The comparison of the 4 zenith angle distributions
with the predictions of the no oscillations MC shows a  disagreement
at low energies, 
while the agreement is restored at the higher neutrino energies.
The distribution of the ratios $R = (Data/MC_{no~osc})$ obtained by this
analysis is plotted in Fig. 2 in the lower part as a function of
$log_{10}(L/E_\nu)$ \cite{macro7}. \par

{\bf{Low energy muons}.} An analysis was done also for the lower energy samples. 
These low energy muons come from neutrinos which have an average energy 
of $\sim 4$ GeV \cite{macro8}. 
The \textit{Internal Upgoing} (IU) muons come from $ \nu _{\mu } $
interactions in the lower apparatus; for these events two scintillation counters
are intercepted; the T.O.F. is applied to identify upward going
muons. 

\begin{figure}[t]
\begin{center}
\mbox{
\includegraphics[width=6.8cm]{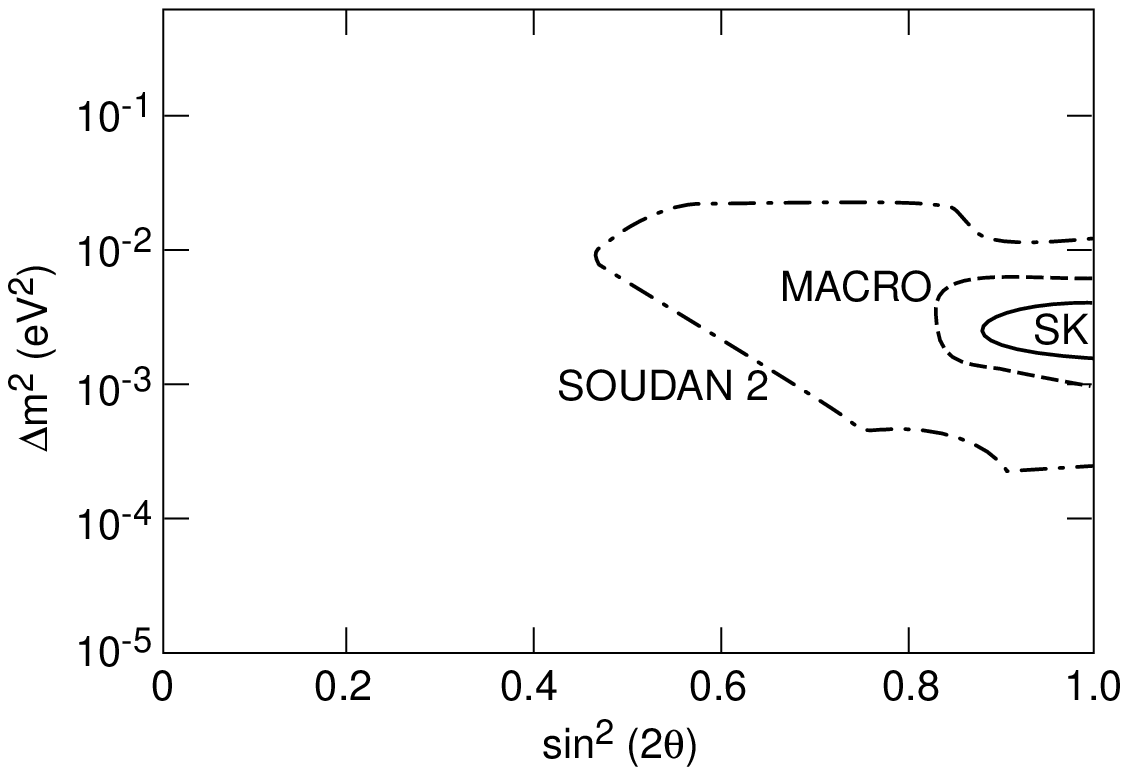}
\hspace{0.2cm} 
\includegraphics[width=6.1cm]{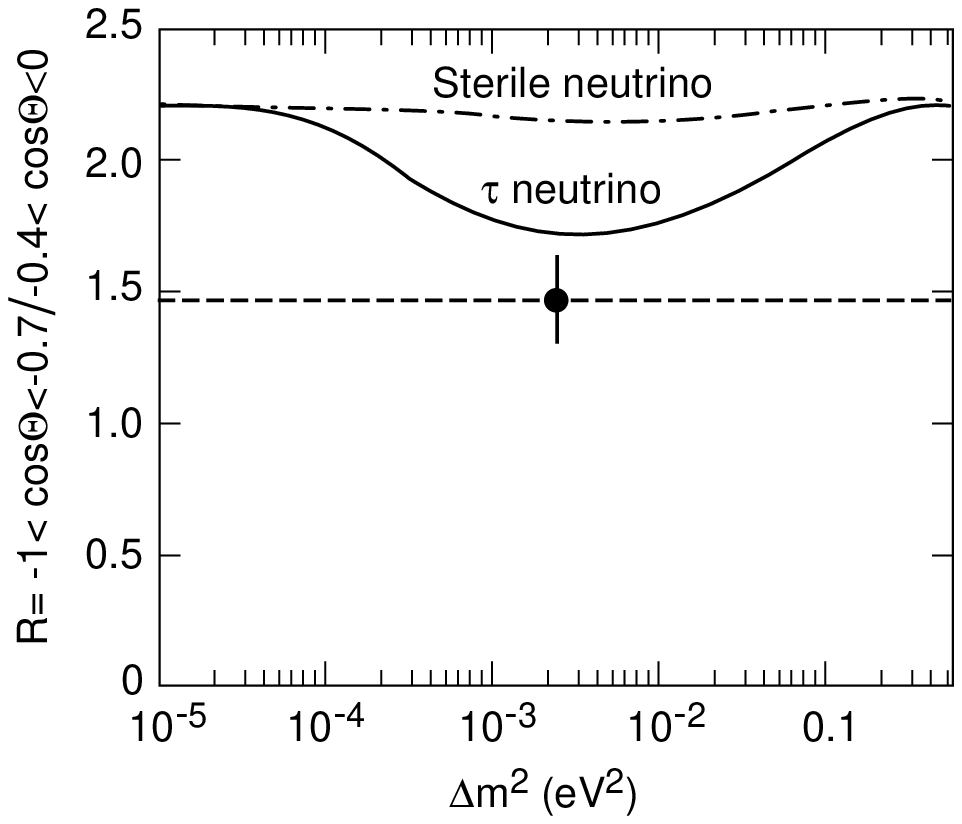}}
\vspace{-0.5cm} 
\includegraphics[width=8cm]{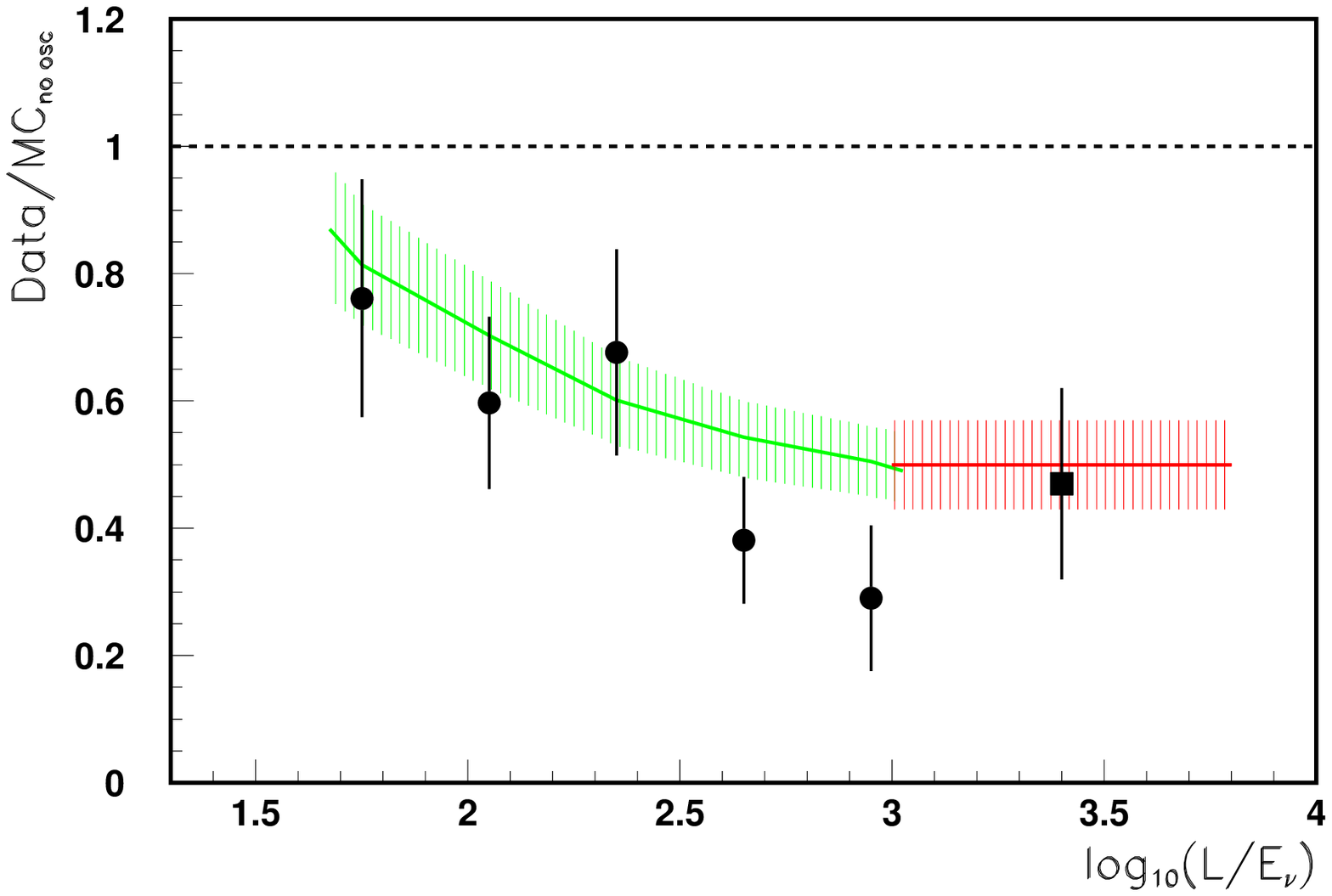}
\end{center}
\caption{\label{fig2} 
\small {\bf{Upper left:}} 90 $\%$ c.l. allowed regions for $\nu_{\mu} \rightarrow \nu_{\tau}$ 
oscillations from up throughgoing muons and comparison with the Soudan 2 and SK allowed regions. 
{\bf{Upper right:}} ratio of events with $-1< cos \theta < -0.7$ to events with
$-0.4<cos\theta<0$ vs $\Delta m^2_{23}$ for maximal mixing. 
The black point is the measured value, the solid line is the
prediction for $ \nu_\mu \rightarrow \nu_\tau $ oscillations, the dash-dotted
line is the prediction for  $ \nu_\mu \rightarrow \nu_{sterile} $ oscillations. 
{\bf{Bottom:}} Data/MC vs $L/E_{\nu}$ for up througoing muons (black circles) and for 
semicontained up-$\mu$ (black square). 
The muon energy was estimated by MCS and $E_{\nu}$ by MC methods. 
The shaded region represents the uncertainty in the MC
prediction assuming $\sin^2 2 \theta=1$ and $\Delta m^2=0.0025$ eV$^2$.
The horizontal  dashed line at Data/MC=1 is the expectation for no
oscillations. } 
\end{figure}

The \textit{upgoing stopping muons} (UGS) are due to external $ \nu _{\mu } $
interactions yielding upgoing muons stopping in the detector. 
The \textit{internal downgoing muons} (ID) are due to $ \nu _{\mu } $-induced
downgoing muon tracks with vertex in the lower MACRO, see Fig. 1 on the left.
A summary table of the number of events is presented with
MC predictions in Table \ref{tab:macro}. 
The data show a uniform deficit for the whole angular distribution with respect to 
predictions, $ \sim 50\% $ for IU, 75\% for ID + UGS; there is a good agreement with the predictions
based on neutrino oscillations with the parameters obtained from the
up throughgoing muons. 
The average value of the double ratio \( R=(Data/MC)_{IU}/(Data/MC)_{ID+UGS} \)
over the measured zenith angle range is \( R\simeq 0.77 \pm 0.07 \); the
error includes statistical and theoretical uncertainties; 
\( R=1 \) is expected in case of no oscillations. \par
In conclusion, MACRO atmospheric neutrino flux analysis gives strong evidence to the  
$\nu_\mu \rightarrow \nu_\tau$ oscillation hypothesis, with 
$\Delta m^2 = 2.5 \times 10^{-3}$ eV$^2$ and maximal mixing.


\begin{table}
{\small \centering \begin{tabular}{cccc}
\hline
                & Events    & MC-No oscillations  & $R = (Data/MC_{no osc}) $\\
                &           &                     &                        \\ \hline
Up throughgoing & $809$ & $1122 \pm 191$ & $(0.721 \pm 0.026_{stat}\pm 0.043_{sys}\pm 0.123_{th} )  $
\\ \hline
Internal Up     & $154$ & $285 \pm 28_{sys}\pm 71_{th}$ & $(0.54 \pm
0.04_{stat} \pm 0.05_{sys}
\pm 0.13_{th} )$\\ \hline
Up Stop +  In Down &  262 & $ 375 \pm 37_{sys} \pm 94_{th} $ & $( 0.70 \pm
0.04_{stat}\pm 0.07_{sys} \pm 0.17_{th})$\\ \hline
\end{tabular}\par}

\caption{{\small Summary of MACRO $\nu _{\mu}\rightarrow \mu$ events in
$-1<cos \theta< 0$ (after background subtraction) for oscillation studies.
For each topology (see Fig. 1 on the left) the number of measured events, the MC
prediction for no-oscillations and the ratio (Data/MC-no osc) are given 
\cite{macro1}, \cite{macro2}, \cite{macro8}.
}}

\label{tab:macro}
\end{table}

\vspace{1cm}
I thank the members of the MACRO collaboration, in particular G. Giacomelli and M. Spurio.

\end{document}